\documentclass[a4paper]{jpconf}
\usepackage{graphicx}
\begin{document}
\title{Infrared Problem in Cold Atom Quantum Physisorption on 2D Materials}
\author{Sanghita Sengupta and Dennis P Clougherty}

\address{Department of Physics, University of Vermont, Burlington, Vermont 05405, USA}

\ead{sanghita.sengupta@uvm.edu}
\ead{dennis.clougherty@uvm.edu}

\begin{abstract}
Results from four different approximations to the phonon-assisted quantum adsorption rate for cold atoms on a 2D material are compared and contrasted: (1) a loop expansion (LE) based on the atom-phonon coupling, (2) non-crossing approximation (NCA), (3) independent boson model approximation (IBMA), and (4) a leading-order soft-phonon resummation method (SPR).  We conclude that, of the four approximations considered, only the SPR method gives a divergence-free result in the large membrane regime at finite temperature.  The other three methods give an adsorption rate that diverges in the limit of an infinite surface.  
\end{abstract}

\section{Introduction}
Current efforts to develop cold atom-based quantum technologies such as precision sensors and quantum information processing have motivated our theoretical interest in developing a fundamental understanding of how cold atoms interact with surfaces.  In the case of atomic adsorption on 2D materials such as suspended samples of graphene, there has been some controversy \cite{dpc2014}; one theory \cite{DPC2013} predicts a  suppression of adsorption for cold atoms, while a numerical calculation has argued for an enhancement of the adsorption rate in comparison to the rate on graphite \cite{LJ2011}.  

% unitarity vs spectral sum rule
We present theoretical results of our studies of the adsorption rate of an atom to an elastic membrane at finite temperature which we propose as a model for phonon-assisted physisorption of cold atoms to 2D materials. We show that this model contains a problem rooted in the emission of soft phonons.  Within the standard perturbation expansion, soft-phonon processes can give large contributions to the adsorption rate that become divergent in the limit of infinite membrane size.  Thus, the perturbation series for large, but finite-sized, membranes can be divergent, and numerical calculations based on the use of a finite-order perturbation expansion of the adsorption rate can be inaccurate for sufficiently large surfaces.

\section{Model}
The 2D material is modeled as an elastic membrane of uniform mass density under tension.  This model is inspired by experiments \cite{mceuen08} that prepare suspended graphene samples by the mechanical deposition of graphene on pitted substrates.  The graphene sample partially adheres to the pit walls of the substrate, consequently straining the sample while it is suspended over the pit.  The strained graphene sample, fixed to the pit walls, is thus viewed as a circular elastic membrane with constant mass density,  and finite radius that is clamped on the edge and is under tension.  Energy can be exchanged between the cold atom and surface by the excitation or deexcitation of vibrational modes of the clamped membrane.  It seems likely that this model could be applied to a variety of 2D materials beyond graphene.  Thus, we view the mass density $\sigma$, transverse speed of sound $v_s$, the radius $a$, and the Debye frequency $\omega_D$ (the high-frequency cutoff of the membrane), as tunable parameters of the model.

The Hamiltonian for small transverse displacements $u({\bf r})$ of the elastic membrane is taken to be 
\begin{equation}
\label{hammem}
H_{ph}=\int  (\frac{1}{2\sigma} \Pi^{2}+\frac{1}{2}\sigma v_s^2 |\nabla u|^{2}) d^2 r
\end{equation}
where $\Pi$ is the canonical momentum density.  This system has been previously quantized \cite{DPC2013}, and the transverse displacement field can expressed in second quantized form
\begin{equation}
\label{u-field}
u({\bf r})=\sum_{m=-\infty}^\infty \sum_{n=1}^\infty \sqrt{\hbar\over 2\sigma \omega_{mn}} \rho_{mn}({\bf r}) (a_{mn}+a^\dagger_{\bar m n})
\end{equation}
where $\rho_{mn}({\bf r})$ are the (normalized) normal modes of transverse vibration for the membrane, and  $\omega_{mn}$ are the corresponding vibrational frequencies for a mode labeled by radial quantum number $n$ and angular quantum number $m$. $a^\dagger_{mn}$ ($a_{mn}$) is the creation (annihilation) operator for a phonon in the mode $(m, n)$.

At long distances, cold atoms interact with the 2D material through a weakly attractive van der Waals (vdW) potential.  Cold atoms are repelled from the surface at short distances because of electron-electron interactions.  A shallow, attractive physisorption well forms with a minimum at a distance of a few $\AA$ from the surface \cite{bonfanti18} in the case of hydrogen adsorption on suspended graphene.  At this distance from the surface, the corrugation of the potential due to graphene's atomic lattice is a small perturbation on the physisorption potential, and we can safely ignore corrugation effects.  

For the static surface, we take potential $V_0(Z)$ with continuous axial symmetry for a cold atom located a distance $Z$ above the plane of the surface's clamped edge. We consider the case of cold atoms initially prepared in a cylindrically symmetric beam directed at normal incidence toward the center of the surface.  Adsorption states in the symmetric potential $V_0(Z)$ can be labeled by angular quantum numbers $m$.  

The atom-phonon interaction for an elastic membrane  has the approximate form \cite{DPC2013}
\begin{equation}
\label{atom-phonon}
H_i= \sum_{n} v(Z) (a_{0n}+a^\dagger_{0 n})
\end{equation}
Thus, an axial beam of atoms at normal incidence will predominantly interact with symmetric ($m=0$) modes.  The angular quantum number $m=0$ will be  dropped for clarity in what follows.

\section{Adsorption Rate: first-order transitions}
Adsorption via the emission of a single phonon is analogous to a capture process \cite{DPC1992, havey2006} of the form $A\to A_b+\phi$ where $A$ is the free atom in the gas phase, $A_b$ is the atom bound to the membrane, and $\phi$ is the phonon emitted during adsorption. This process is
energetically possible for low-energy atoms when the binding energy of the atom to the membrane $E_b$ is below the maximum energy of a single phonon $\omega_D$.  (We use natural units where $\hbar=k_B=1$.)  

To elucidate the fundamentals of the process, we analyze the first-order phonon-assisted adsorption rate using Fermi's ``golden rule.''  We take the thermal average of the initial phonon distribution and calculate the adsorption rate $\Gamma^{(0)}$ via single phonon emission.  We consider the target membrane to be initially in thermal equilibrium at temperature $T$.    We obtain  an adsorption rate of the following form
\begin{eqnarray}\label{GR}
%\begin{split}
 \Gamma^{(0)} &=& 2\pi \sum_{f, i} |\langle f|H_{i}|i\rangle|^{2} \delta (E_{f}-E_{i}) p_i \nonumber\\
 &=& 2\pi \sum_{q} |\langle B, n_{q}+1| v (a_{q}+a^\dagger_{q})|K, n_{q}\rangle|^{2}\delta(E_s-\omega_{q})p(n_q)\nonumber\\
 \Gamma^{(0)}  &=&{2\pi g_{kb}^{2}\rho} ({n}(E_{s}) +1) 
%\end{split}
\end{eqnarray}
where $p(n_q)$ is the probability at temperature $T$ of initially having $n_q$ phonons in the qth mode, $|B\rangle$ is the bound state of the atom in the static potential $V_0$ with energy $-E_b$, $|K\rangle$ is the initial state of the atom with energy $E$, and $\rho$ is the (partial) vibrational density of states for circularly symmetric modes.  The energy $E_s\equiv E+E_b$ is the energy transferred from the atom to the membrane during the adsorption process.  The coupling constant $g_{kb}$ is the transitional matrix element for the atom
\begin{equation}
g_{kb}\equiv\langle B|v|K\rangle
\end{equation}

We observe that the thermally-averaged phonon occupancy of the mode with phonon energy $E_s$, $n(E_s)$,  appears in the adsorption rate.  The adsorption rate $\Gamma^{(0)}$ is enhanced relative to the zero temperature case by a factor of ${n}(E_{s}) +1$.  This is a simple consequence of the enhanced probability of phonon creation in a solid at non-zero temperature where phonons are initially present.  

While the first-order rate $\Gamma^{(0)}$ is finite, we find higher order contributions tend to diverge in the limit of infinite membrane size.  This calls to question the convergence of the perturbation series for large membranes.  It also casts doubt on the validity of approximations that merely truncate the series at leading order, a common practice in the computational literature.

\section{Adsorption Rate: higher-order transitions}
%\subsection{Atom self-energy}
An efficient way to calculate the contributions of the adsorption rate from multiphonon emission is to calculate the atom self-energy $\Sigma_{kk}(E)$ using diagrammatic perturbation theory.  The imaginary part of the atom self-energy is proportional to the decay rate \cite{Sengupta2017}.  For cold atoms in the gas phase where the atom's energy $E$ is close to zero, the only decay channel available for our model is adsorption to the membrane.  The phase space available for inelastic scattering where the atom remains in the gas phase tends to zero as $E\to 0$ \cite{DPC1992}.  Thus, we relate the low-energy adsorption rate $\Gamma$ to the atom self-energy in the gas phase $\Sigma_{kk}(E)$ by
\begin{equation}
\Gamma=-2 Z {\rm Im}\ \Sigma_{kk}(E)
\end{equation}
where the renormalization factor $Z$ is given by 
\begin{equation}\label{RF}
Z = \bigg[1- {\rm Re}\ \bigg({\partial\Sigma_{kk}(E)\over\partial E}\bigg)\bigg]^{-1}
\end{equation}

The one-loop self-energy $\Sigma_{kk}^{(1)}$ reduces to the lowest-order golden rule result in Eq.~\ref{GR} for sufficiently weak $g_{kb}$ where $Z\approx 1$.  We obtain a one-loop atom self-energy of the form
\begin{equation}
\Sigma_{kk}^{(1)}=g_{kb}^2\sum_q \bigg({n(\omega_q)+1\over E+E_b-\omega_q+i\eta}+ 
{n(\omega_q)\over E+E_b+\omega_q+i\eta} \bigg)
\label{selfenergy}
\end{equation}

Thus,
\begin{equation}
{\rm Im}\ \Sigma_{kk}^{(1)}=-\pi \rho g_{kb}^2 \int_\epsilon^{\omega_D} d\omega (n(\omega)+1) \delta(E+E_b-\omega)
=-\pi \rho g_{kb}^2 (n(E_s)+1)
\end{equation}
and 
\begin{equation}
\Gamma^{(1)}=2 \pi \rho g_{kb}^2 (n(E_s)+1)
\end{equation}
for $Z=1$. 

For a membrane of finite size, its vibrational spectrum is bounded below by a non-vanishing frequency $\epsilon$.  As the size of the membrane increases, this lower bound to the spectrum $\epsilon$ tends to zero frequency.  We will see that $\epsilon$ serves as a regulator of infrared divergent contributions to the atom self-energy.  The Debye frequency $\omega_D$ serves as a high-frequency cutoff to the vibrational spectrum.  $\rho$ is the (partial) density of vibrational states with circular symmetry.

We find that the loop expansion of the self-energy contains contributions that diverge in this limit of $\epsilon\to 0$.   For a membrane at non-zero temperature, this infrared divergence appears already at the one-loop level.
For suitably large membranes where the vibrational frequencies form a dense set, we approximate the real part of the one-loop self-energy $\Sigma_{kk}^{(1)}$ in Eq.~\ref{selfenergy} for $\epsilon\ll T\ll E_b$ by integration to get
\begin{eqnarray}
{\rm Re}\ \Sigma_{kk}^{(1)}&=&\rho g_{kb}^2\ {\cal P}\int_\epsilon^{\omega_D} d\omega \bigg({n(\omega)+1\over E+E_b-\omega} +{n(\omega)\over E+E_b+\omega}\bigg)\nonumber\\
&\approx& {2\rho g_{kb}^2 T\over E_s} \ln\frac{T}{\epsilon}
\label{realSE}
\end{eqnarray}
Thus, the one-loop contribution to the atom self-energy is logarithmically divergent as $\epsilon\to 0$.

At the two-loop level, both real and imaginary parts of the atom self-energy are divergent in this large membrane limit.  The degree of divergence increases with increasing order in the loop expansion of the self-energy (see Table~\ref{table:loop}).  Thus, approximating the adsorption rate by truncating the loop expansion of the atom self-energy to finite order is invalid as $\epsilon$ tends to zero.  

\begin{center}
\begin{table}[h]
\caption{\label{table:loop} The real and imaginary parts of the atom self-energy $\Sigma_{kk}$ at the one and two-loop levels for $T\gg E_b\gg E$.  In the limit of $\epsilon\to 0$ (infinite membrane size), ${\rm Im}\ \Sigma_{kk}^{(1)}$ is finite, while ${\rm Re}\ \Sigma_{kk}^{(1)}$ diverges logarithmically. Both real and imaginary parts of the two-loop self-energy $\Sigma_{kk}^{(2)}$ diverge as $\epsilon\to 0$. The coupling constant $g$ is defined by $g \equiv \rho g_{kb}g_{bb}$. \\}
\centering
 %   {\rowcolors{1}{red!50!red!50}{green!70!red!40}
  \begin{tabular}{  c  c  c     }
\mr
    Feynman Diagram & Re $\Sigma_{kk}$ & Im $\Sigma_{kk}$ \\ 
\mr
    1-loop  &  \ $\frac{2\rho g_{kb}^{2}T}{E_{b}}\ln\frac{E_{b}}{\epsilon}$ &  $-\frac{\pi \rho g_{kb}^{2}T}{E_{b}} $ \\ 
    \\
    2-loop & $ \frac{2 g^{2}T^{2}}{E_{b}^{3}}\ln^2\big(\frac{E_{b}}{\epsilon}\big)$ & $-\frac{ g^{2}T^{2}}{2\pi E_{b}^{2}}\frac{1}{\epsilon}$\\
\mr
  \end{tabular}
 % }

\end{table}
\end{center}

\section{Resummation methods}

The infrared divergences in the loop expansion of the atom self-energy were previously discussed both for a zero temperature \cite{DPC2017, Sengupta2016} and a finite temperature \cite{dpcSPR2017} membrane.  At $T=0$, we know from the Kinoshita-Lee-Nauenberg (KLN) theorem that these infrared divergences are an artifact of the perturbation expansion and that the physical adsorption rate is free of such divergences.  While the KLN theorem has not been generalized to $T\ne 0$ formally, it seems very likely that it also applies at finite temperature.
There are field theoretic examples at finite temperature \cite{ALTHERR1990} where the explicit cancellation of the infrared singularities has been demonstrated to finite order.  Motivated by these examples, we consider in what follows three resummation methods that were previously successful in removing the infrared singularities \cite{Sengupta2017} in the adsorption rate at $T=0$.

\subsection{Independent boson model}
When we truncate the atom's Hilbert space to two states, our physisorption model reduces to a two-channel generalization of the independent boson model (IBM).  Consider the two essential eigenstates of $H_a$ in describing the physisorption dynamics: the initial atomic state  $|K\rangle$ at low (positive) energy and the final atomic state, a bound state $|B\rangle$ with binding energy $E_b$.  The Hamiltonian in this truncated state space \cite{DPC2013} has the form $H=H_a+H_p+H_{i1}+H_{i2}$ where 
\begin{eqnarray}
H_a&=& E_k c^\dagger_k c_k -E_b b^\dagger b\\
H_p&=& \sum_n \omega_n a^\dagger_n a_n\\
H_{i1}&=&-g_{kb}(c_k^\dagger b+b^\dagger c_k) \sum_n  (a^\dagger_n +a_n)\\
H_{i2}&=&-g_{bb} b^\dagger b \sum_n  (a^\dagger_n +a_n)
\end{eqnarray} 
Here, the second coupling constant $g_{bb}\equiv\langle B|v|B\rangle$. $b^\dagger$ creates the bound atom, and $c_k^\dagger$ creates the atom in the state $|K\rangle$ with energy $E_k$. 

\begin{figure}[h]
\begin{center}
\includegraphics[width=0.5\columnwidth]{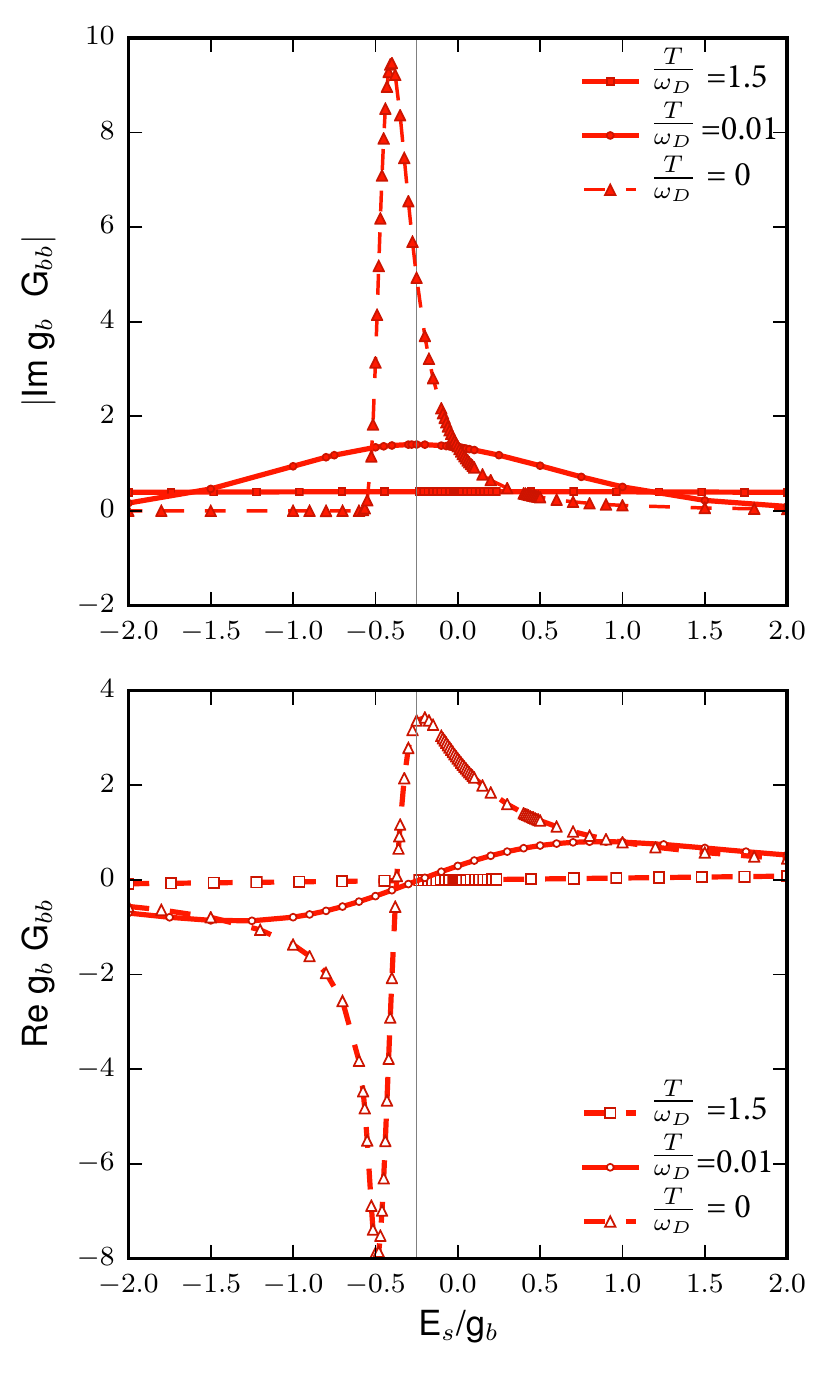} 
\end{center}
\caption{\label{fig:IBMpropagator} Real and imaginary parts of the bound atom Green function $G_{bb}^{\rm IBM}$ vs energy $E_s$ for  temperatures $T= 1.5\ \omega_{D}$, $T=0.01\ \omega_{D}$ and $T=0$. The infrared cutoff has constant value $\epsilon=0.01 g_b$ where $g_b\equiv\rho g_{bb}^2$.  ${\rm Im}\ G_{bb}^{\rm IBM}$ is phonon-broadened at $T=0$, and the broadening increases with increasing $T$.} 
\end{figure}

For low-energy atoms, $g_{kb}$ is reduced in strength by quantum reflection \cite{DPC1992} and becomes weaker with decreasing energy.  Thus, we focus on the regime where  $g_{kb}\ll g_{bb}$, corresponding to weak coupling between the IBM channels, and we calculate the adsorption rate to lowest order in $g_{kb}$.  The effect of $g_{bb}$ is included to all orders by using the exact atom Green function of the IBM $G_{bb}^{\rm IBM}$ (see Fig.~\ref{fig:IBMpropagator}) in the expression for the one-loop atom self-energy.  We refer to this scheme as the independent boson model approximation \cite{Sengupta2017, Sengupta2016} (IBMA).  Thus,
\begin{equation}
\Sigma_{kk}^{(\rm IBM)}(E)= g_{kb}^{2} \sum_{q}\bigg(n_{q} G_{bb}^{\rm IBM} (E+\omega_{q})+ (n_{q} +1) G_{bb}^{\rm IBM} (E-\omega_{q})\bigg)\\
\label{IBMA}
\end{equation}
% asymptotic rate calculation

\begin{equation}
\Gamma^{(\rm IBM)}(E)= 2\pi g_{kb}^{2}\rho Z \int_\epsilon^{\omega_D} d\omega\bigg(n({\omega})\rho_b^{\rm IBM} (E+\omega)+ (n({\omega}) +1) \rho_b^{\rm IBM} (E-\omega)\bigg)\\
\label{ibm-rate}
\end{equation}
where $\rho_b^{\rm IBM} (E)\equiv -\frac{1}{\pi} {\rm Im}\ G_{bb}^{\rm IBM} (E)$, the phonon-broadened bound state spectral density.  From the exact IBM Green function, we obtain a simple Gaussian form for the bound state spectral density in the asymptotic regime of large temperature by using the method of Laplace.  Explicitly, we find
\begin{equation}
\rho_b^{\rm IBM} (E)={\exp\big({-{(E_s+\Delta)^2}/{4\alpha}}\big)\over \sqrt{4\pi\alpha}}
\end{equation}
where $\Delta=\rho g_{bb}^2  \ln\frac{\omega_D}{\epsilon}$ and $\alpha =T\Delta$.  $\Delta$ provides a shift to the atom binding energy that results from the atom-phonon coupling for the bound atom.  

As $\epsilon\to 0$, the relative adsorption rate ${\Gamma^{(\rm IBM)}/ \Gamma^{(0)}}$ behaves asymptotically as
\begin{equation}
{\Gamma^{(\rm IBM)}\over \Gamma^{(0)}}= \sqrt{Z^2 \rho g_{bb}^2 \over \pi  T} \bigg(\ln\frac{\omega_D}{\epsilon}\bigg)^{\frac{3}{2}}, \ \ \ \epsilon\to 0
\label{ibm-div}
\end{equation}
where $Z^{-1}\sim (1-g_{kb}^2/g_{bb}^2)$ for large T.
We conclude that, in contrast to the zero temperature case, a weak (logarithmic) infrared singularity remains in IBMA for finite temperature.

\subsection{Non-crossing approximation}
The non-crossing approximation (NCA) for the atom self-energy contains an infinite summation over the class of nested Feynman diagrams.  This result is formally equivalent to solving the NCA equations \cite{Sengupta2017}, a system of coupled nonlinear integral equations for the self-energy; namely, 
\begin{eqnarray}
    \Sigma_{bb}(E) &=& g_{bb}^{2}\sum_{q}\bigg[\frac{n_{q}}{E+E_{b}+\omega_{q}-\Sigma_{bb}(E+\omega_{q})} + \frac{n_{q}+1}{E+E_{b}-\omega_{q}-\Sigma_{bb}(E-\omega_{q})}\bigg] \nonumber\\
    &+&  g_{kb}^{2}\sum_{q}\bigg[\frac{n_{q}}{E-E_{k}+\omega_{q}-\Sigma_{kk}(E+\omega_{q})} + \frac{n_{q}+1}{E-E_{k}-\omega_{q}-\Sigma_{kk}(E-\omega_{q})}\bigg] 
\end{eqnarray}    
\begin{eqnarray}
    \Sigma_{kk}(E) &=& g_{kb}^{2}\sum_{q}\bigg[\frac{n_{q}}{E+E_{b}+\omega_{q}-\Sigma_{bb}(E+\omega_{q})} + \frac{n_{q}+1}{E+E_{b}-\omega_{q}-\Sigma_{bb}(E-\omega_{q})}\bigg] 
\end{eqnarray}

\begin{figure}[h]
\begin{center}
\includegraphics[width=0.72\columnwidth]{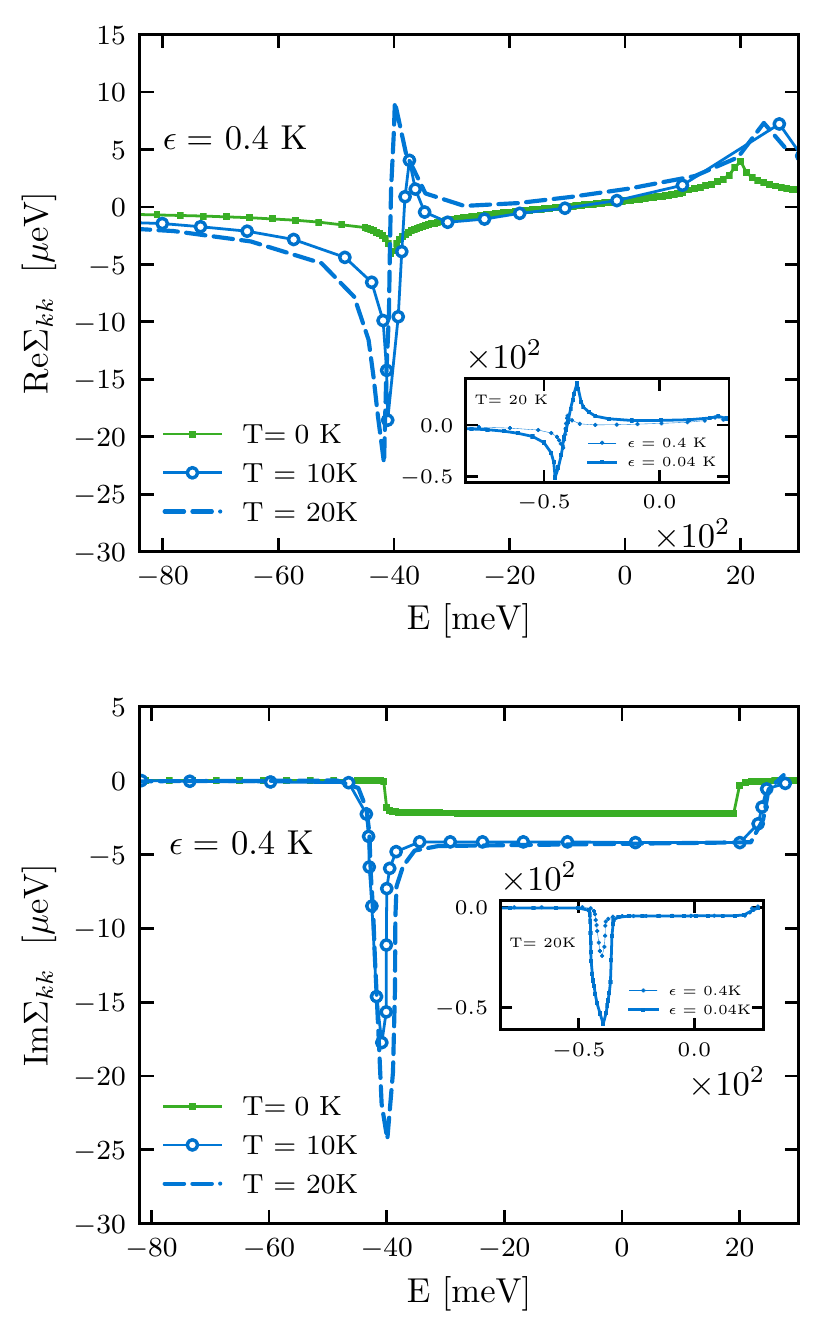} 
\end{center}
\caption{\label{fig:TSkk} Top: Real part of the NCA atom self-energy $\Sigma_{kk} (E)$ vs atom energy $E$ for membrane temperatures $T=0, 10$ and 20 K and $\epsilon= 0.4$ K.  Bottom: Imaginary part of the NCA atom self-energy $\Sigma_{kk} (E)$ vs atom energy $E$ for membrane temperatures $T=0, 10$ and 20 K and $\epsilon= 0.4$ K.  Inset plots: NCA atom self-energy $\Sigma_{kk} (E)$ vs atom energy $E$ at $T=20$ K for $\epsilon= 0.4$ and $0.04$ K.} 
\end{figure}

The self-consistent NCA atom self-energy is obtained numerically by successive iteration of the coupled NCA equations until convergence is achieved.  Numerical values of the model parameters are given in Table~\ref{table:params}.  Fig.~\ref{fig:TSkk} shows the variation of the real and imaginary parts of the NCA atom self-energy with atom energy $E$ and membrane temperature $T$ at fixed IR cutoff $\epsilon$.  

\begin{center}
\begin{table}[h]
\caption{\label{table:params} Parameter values for atomic hydrogen adsorbing on suspended graphene.}
\centering
 %   {\rowcolors{1}{red!50!red!50}{green!70!red!40}
  \begin{tabular}{  c  c  c   c  c  c  c}
\mr
    $g_{bb}^2\rho$ (meV)& $g_{kb}^2\rho$ ($\mu$eV)& $\omega_D$ (meV)&$T$ (K)&$E_b$ (meV)&$\epsilon$ (K)&$v_s$ (m/s)\\ 
\mr
   0.06  & 0.5-10 &  65& 0-20& 40&0.01-0.5& $6.64\times 10^3$\\ 
\mr
  \end{tabular}
 % }

\end{table}
\end{center}

\begin{figure}[h]
\begin{center}
\includegraphics[width=0.65\columnwidth]{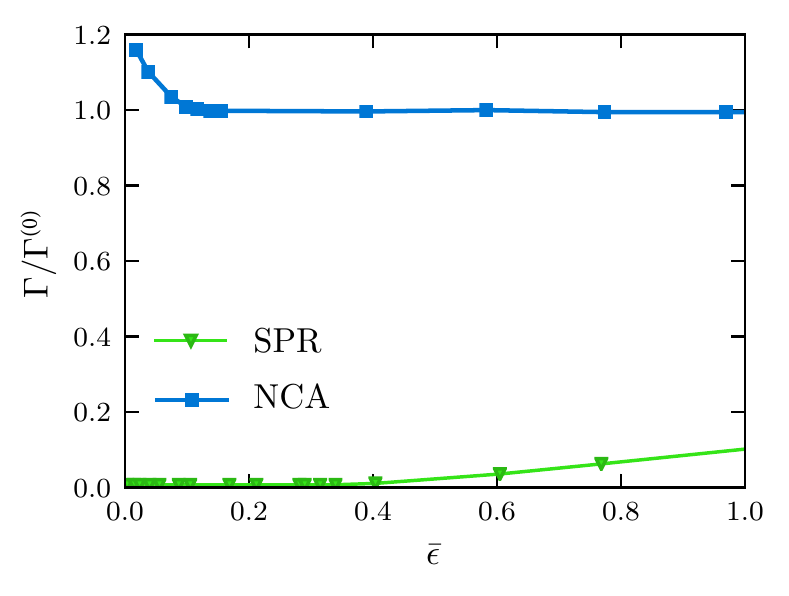} 
\end{center}
\caption{\label{fig:SPR} Normalized adsorption rate  $\Gamma/\Gamma^{(0)}$ vs (scaled) infrared cutoff $\bar\epsilon\equiv\epsilon/g_{bb}$ for atom energy $E=0$ at (scaled) temperature $T=15 g_{bb}$ for SPR and NCA.  In the regime of small IR cutoff, the SPR adsorption rate is suppressed, while the NCA adsorption rate grows relative to $\Gamma^{(0)}$ as $\bar\epsilon\to 0$.} 
\end{figure}

For $\epsilon\ll 0.1 g_{bb}$, we find numerically that $\Gamma^{\rm NCA}/\Gamma^{(0)}$ tends to diverge with decreasing $\epsilon$  (see Fig.~\ref{fig:SPR}),  We attribute this to the zero frequency singularity in the phonon distribution function at finite temperature $n(\omega)$. Thus, the infrared divergence problem persists within NCA for finite temperature.

%approximate solution showing IR divergence

\subsection{Soft-phonon resummation}
In the soft-phonon resummation (SPR) method \cite{dpcSPR2017}, we take $H_{i1}$ as the sole perturbation, recognizing that the remainder of $H$ contains the exactly solvable independent boson model.  Thus, the unperturbed Hamiltonian, $H_0\equiv H_a+H_p+H_{i2}$, contains atom-phonon interactions for the bound atom.

We now  consider transitions under $H_{i1}$ to a final state where the atom is bound to the membrane with the emission of $n$ phonons: a hard phonon whose energy of the order of the binding energy $E_b$, and $(n-1)$ soft phonons, each with energies of the order of the IR cutoff $\epsilon$.    We calculate the leading order of this transition amplitude as $\epsilon\to 0$ for $n$-phonon emission and sum the resulting transition rates $\Gamma_n$ to obtain the total adsorption rate at zero temperature \cite{DPC2017}.  

Because of the inclusion of atom-phonon interactions in $H_0$, the unperturbed ground state is a polaronic state consisting of the atom bound to the membrane, accompanied by a cloud of phonons.  At zero temperature, the phonon portion of the ground state is a phonon coherent state generated by applying a unitary operator that displaces the initial phonon vacuum \cite{DPC2017}.  

At zero temperature, the adsorption rate via the emission of $n$ phonons $\Gamma_n$ is vanishingly small for a suitably large membrane \cite{DPC2017}; however, the total adsorption rate obtained by summing over the emission of all possible soft phonons is still suppressed but can be a non-negligible fraction of the golden rule result $\Gamma^{(0)}$, with
\begin{equation}
\sum_{n=1}^\infty \Gamma_n={\cal R}\ \Gamma^{(0)}
\end{equation}
where the total reduction factor ${\cal R}=(1+{\Delta/ E_s})^{-2}$, vanishes only logarithmically fast with decreasing IR cutoff $\epsilon$.

For $T\ne 0$, we compute the transition rate via the emission of $n$ phonons (one hard and $n-1$ soft phonons), taking an initial phonon distribution $\{n_q\}$. The rate is then thermally averaged over the initial phonon distribution.  The resulting rate contains a phonon reduction factor that is both cutoff and temperature dependent, vanishing exponentially fast as $\epsilon\to 0$ \cite{dpcSPR2017}. In contrast to the zero temperature case, the total rate at finite temperature, obtained by summation over all possible soft-phonon emissions, vanishes exponentially fast as $\epsilon\to 0$ (see Fig.~\ref{fig:SPR}).

We conclude that using phonon coherent states as a basis for the final states gives transition matrix elements that are: (1) IR divergence-free, and (2) vanish exponentially fast as $\epsilon\to 0$.   Thus, at finite temperature, the adsorption rate is exponentially suppressed from $\Gamma^{(0)}$ in the regime of small IR cutoff.  It is the improved treatment of the final phonon states as generalized coherent states that cures the infrared problem in the quantum adsorption on membranes.

%======================================================

\section*{References}
\bibliographystyle{iopart-num}
\bibliography{QFT}

\end{document}